\documentclass[eqsecnum,showkeys,showpacs,nofootinbib,aps,epsfig]{revtex4}
\renewcommand{\theequation}{\arabic{equation}}
\usepackage{graphicx}
\def\bea{\begin{eqnarray}}
\def\eea{\end{eqnarray}}

\newcommand{\nn}{\nonumber}

\def\beq{\begin{equation}}
\def\eeq{\end{equation}}

\def\pa{\partial}

\newbox\pippobox
\def\d{{\rm d}}
\def\btheta{\bar{\theta}}
\def\bpsi{\bar{\psi}}
\def\bdelta{\bar{\delta}}
\def\ol{\overline}
\def\Dt{D_{t}}

\begin{document}
\title{Supersymmetric Monopole Quantum Mechanics on Sphere}
\author{Soon-Tae Hong}
\email{soonhong@ewha.ac.kr} \affiliation{Department of Science
Education, Ewha Womans University, Seoul 120-750
Korea}
\author{Joohan Lee}
\email{joohan@kerr.uos.ac.kr} \affiliation{Department of Physics,
University of Seoul, Seoul 130-743 Korea}
\author{Tae Hoon Lee}\email{thlee@ssu.ac.kr}
\affiliation{Department of Physics, Soongsil University, Seoul
156-743 Korea}
\author{Phillial Oh}
\email{ploh@newton.skku.ac.kr} \affiliation{Department of Physics
and Institute of Basic Science, Sungkyunkwan University, Suwon
440-746 Korea}
\date{\today}%
\begin{abstract}
We study $N=2$ supersymmetric quantum mechanics of a charged
particle on sphere in the background of Dirac magnetic monopole.
We adopt $CP(1)$ model approach in which the monopole interaction
is free of singularity. It turns out that this approach admits a
compact $N=2$ superspace formulation. In order to exploit manifest
$U(1)$ covariance in the superspace formalism, we introduce a
gauged chiral superfield which is annihilated by the gauge
covariant superderivative instead of the usual superderivative. We
carry out the Dirac quantization of the resulting system and
compute the quantum mechanical spectrum. We obtain the condition
for the spontaneous breaking of supersymmetry explicitly in terms
of the monopole charge and a parameter which characterizes the
operator ordering ambiguity. We find that the supersymmetry is
spontaneously broken unless a certain combination of these
quantities satisfies some quantization condition.
\end{abstract}
\pacs{11.30.Pb, 11.30.Qc, 14.80.Hv} \keywords{supersymmetric quantum
mechanics; magnetic monopole; operator ordering; supersymmetry
breaking } \preprint{hep-th/yymmnn} \maketitle

\section{Introduction}
\setcounter{equation}{0}
\renewcommand{\theequation}{\arabic{section}.\arabic{equation}}

Quantum mechanics in the background of Dirac magnetic
monopole~\cite{dira} exhibits many interesting features such as
quantization of the electric charge, modified orbital angular
momentum and hidden conformal symmetries associated with the time
reparametrization invariance~\cite{cole, jack}. The supersymmetric
magnetic monopole quantum mechanics has attracted a great deal of
attention recently due to the existence of hidden superconformal
symmetry~\cite{hoke} and in relation with superconformal
mechanics~\cite{leiv}. In this paper, we study $N=2$ supersymmetric
quantum mechanics~\cite{coop} of a charged particle on a sphere in
the background of Dirac magnetic monopole by using the chiral
superfield  formalism.

Our motivation for considering this system is twofold. One is that
the supersymmetric quantum mechanics on general target manifold
(regardless of the presence of magnetic monopole) is interesting in
itself and its study revealed many important aspects of
supersymmetry~\cite{witt}. The other concerns with the number of
supersymmetries allowed when the magnetic monopole interaction is
present. It is well known that the chiral $N=2$ superfield
formulation of supersymmetric quantum mechanics on $S^2$ is possible
because of its K\"ahler structure~\cite{zumi, akul}. However, as far
as the magnetic monopole interaction is concerned, most of the
previous work dealt with $N=1$ superfields in ${\bf R}^3$
\cite{hok}. Later, it was found that this system admits another
supersymmetry and its relation with the constrained dynamics on a
sphere was discussed~\cite{dejo}. More recently, $N=2$
supersymmetric quantum mechanics of a charge-monopole system
confined to $S^2$ has been investigated in terms of unconstrained
variables~\cite{kim}. Utilizing the shape-invariance of the system,
they obtained the energy spectrum and a complete set of energy
eigenstates~\cite{kim} with a particular choice of the operator
ordering~\cite{alfaro}.

In this work, we adopt $CP(1)$ model approach~\cite{adda} where the
dynamical variables take the value on $S^3$, but the dynamics is
reduced to $S^2$ by imposing $U(1)$ gauge symmetry. This has the
merit that the quantum mechanical Lagrangian of magnetic monopole is
free of singularity~\cite{bala} and one does not have to deal with
the multi-valued action~\cite{wu}. It also has the advantage that
the rotational generators are well realized. We find that this
approach allow a compact $N=2$ superfield formulation of the system
including the monopole interaction. In order to exploit the $U(1)$
gauge covariance we introduce gauged chiral superfield which is
annihilated by the gauge covariant superderivative, in which the
time derivative in the usual superderivative is replaced with the
gauge covariant one. This gauge covariant superderivative still
satisfies the  usual important property that it commutes with the
supersymmetry generators (see Eqs (\ref{ddbar}),  (\ref{qandqbar})
and (\ref{nilpo})).

In quantizing the system there appears a parameter associated with
the choice of operator ordering in defining the basic commutation
relations. We study how physical quantities such as energy and
angular momentum depend on this parameter. We obtain the exact
quantum mechanical energy spectrum and discuss the possibility of
spontaneous breaking of supersymmetry in terms of the monopole
charge and the ordering parameter.

Let us briefly recall the bosonic $CP(1)$ model. The Lagrangian is
given by \beq L_{0}= 2\vert D_{t} z_i \vert^2, \eeq with $D_{t} =
\pa_{t}-ia$, where $a$ is the auxiliary field. Impose the condition
$\bar z \cdot z\equiv\sum_{i=1}^{2} \vert z_i \vert^2\ = 1$ with $
z=\left(\begin{array}{c}z_{1}\\z_{2}\end{array}\right)$. Due to the
$U(1)$ invariance, the dynamics is reduced from $S^3$  to $S^2$,
which is the $CP(1)$ model.  Eliminating the auxiliary field $a$ by
using the equation of motion, we obtain \beq L_{1}=2|\dot{z} +
(\dot{\bar{z}}\cdot z ) z|^{2}, \eeq where the overdot denotes the
time derivative. The magnetic monopole background interaction is
given by \beq L_{2}= ig(\bar z\cdot\dot{z}-\dot{\bar{z}}\cdot z),
\eeq where $g$ is the magnetic monopole charge\footnote{Here, we set
the electric charge $e=-1$.}. Observe that the Lagrangian $L_1$ is
invariant under the $U(1)$ gauge symmetry generated by $z\rightarrow
e^{i\Lambda(t)}z$, whereas $L_2$ changes by a total time derivative.
They are free of singularities. The singular Lagrangian emerges
through the Hopf fibration $\vec x = \bar z \vec \sigma z$
\cite{ryde} and the introduction of local coordinates
$z_1=1/\sqrt{1+\vert\xi\vert^2},~ z_2=\xi/\sqrt{1+\vert\xi\vert^2}$.
The stereographic projection $\xi= \tan\frac{\theta}{2} e^{i\phi}$
produces the monopole interaction which is singular along the
negative $z$-axis. The choice $\xi= \cot\frac{\theta}{2} e^{i\phi}$
gives singularity along the positive $z$-axis. Both choices yield
the standard kinetic Lagrangian $L_1= \frac{1}{2}\dot{\vec
x}\cdot\dot {\vec x}$.

This paper is organized as follows.  In section 2, we define gauge
covariant superderivatives and gauged chiral superfield.  We then
construct $N=2$ supersymmetric monopole Lagrangian. In section 3, we
quantize the system via Dirac quantization method and discuss
operator ordering ambiguity.  In section 4, we compute the energy
spectrum and analyze the spontaneous supersymmetry breaking
phenomena.  Section 5 includes summary and discussions.

\section{$N=2$ supersymmetric monopole Lagrangian}
\setcounter{equation}{0}
\renewcommand{\theequation}{\arabic{section}.\arabic{equation}}

We present our $N=2$ supersymmetric Lagrangian in a $U(1)$ covariant
manner. First,  we introduce superspace $(t, \theta, \bar\theta)$
and  define the gauge covariant superderivatives as \bea
D &=& \pa_{\theta}-i\btheta D_{t},\nn\\
\bar{D} &=& \pa_{\btheta}-i\theta D_{t}, \label{ddbar}\eea where
$D_{t}$ denotes  $U(1)$ covariant derivative, \bea\Dt
z &=& \left(\pa_{t}-ia\right)z,\nn\\
\Dt\bar{z} &=& \left(\pa_{t}+ia\right)\bar{z}.\eea for some real
field $a$, which we will specify shortly.  Note that the covariant
superderivatives $D$ and $\bar{D}$ satisfy \beq
D^{2}=\bar{D}^{2}=0,~~~[D,\bar{D}]_+=-2iD_{t}.\eeq  $N=2$ gauged
chiral superfield, $\Phi$, is defined as usual by imposing the
condition $\bar{D}\Phi=0$. Thus, we get \beq \Phi = z + \theta \psi
-i\theta\btheta D_{t}z. \label{phiz}\eeq Supersymmetry generators
are similarly modified to \bea
Q &=& \pa_{\theta}+i\btheta D_{t},\nn\\
\bar{Q} &=& \pa_{\btheta}+i\theta D_{t}, \label{qandqbar}\eea which
satisfy \beq [Q,\bar{Q}]_+=2i\Dt, \label{qqbar} \eeq  and fulfill
the relations \beq
Q^{2}=\bar{Q}^{2}=0,~~~[D,Q]_+=[D,\bar{Q}]_+=[\bar{D},Q]_+=[\bar{D},\bar{Q}]_+=0.
\label{nilpo}\eeq

Transformation rules of the chiral field components are obtained by
applying $Q$ and $\bar{Q}$ to the superfields.  From \bea Q\Phi
&=&\delta z-\theta\delta \psi-i\theta\btheta
\delta(D_{t}z)\nn\\&=&\psi-i\theta\btheta D_{t}\psi,\label{qphi}\eea
we get \beq \delta z = \psi,~~~\delta \psi=0,~~~
\delta(D_{t}z)=\Dt\psi.\eeq Similarly, from \bea
Q\bar{\Phi}&=&\delta \bar{z}+\btheta\delta \bar{\psi}+i\theta\btheta
\delta(D_{t}\bar{z})\nn\\&=&2i\btheta D_{t}\bar{z}, \eea we find
\beq \delta \bar{z} =0,~~~\delta \bpsi= 2iD_{t}\bar{z},~~~
\delta(D_{t}\bar{z})=0. \eeq Note that both of these transformation
rules require for consistency that \beq \delta a=0. \label{dela}\eeq
Calculation for $\bar{Q}$ leads to the similar transformation rules
for the component fields and the consistency condition, $\bar\delta
a=0$. We summarize the supertransformation of the fields,\beq
\begin{array}{llll}
\vspace{0.2cm}
\delta z = \psi, &\delta \bar{z} = 0, &\delta \psi=0, &\delta \bpsi=2i{\Dt\bar z},\\
\bdelta z = 0, &\bdelta \bar{z} = \bar{\psi}, &\bdelta \psi=2i{\Dt
z}, &\bdelta \bpsi=0.
\end{array}\eeq \label{dtz}

Our supersymmetric action is then proposed by \beq L=\int
\d\btheta\d\theta~\left(\frac{1}{2}\ol{D\Phi}\cdot
D\Phi\right)-2ga\eeq with the superconstraint \beq
\bar{\Phi}\cdot\Phi-1=0. \label{susyconst}\eeq This superfield
constraint incorporates the familiar constraints~\cite{adda},
$\bar{z}\cdot z-1=0$, $\bar{z}\cdot\psi=0=\bar{\psi}\cdot z$, and
determines the field $a$, \beq
a=-\frac{i}{2}(\bar{z}\cdot\dot{z}-\dot{\ol{z}}\cdot
z)-\frac{1}{2}\bpsi\cdot\psi. \label{aeq} \eeq It is important to
note that the field $a$ obtained above is indeed invariant under the
supersymmetry transformations to yield Eq. (\ref{dela}).  This is
what makes our whole construction consistent.  After performing the
$\theta$ and $\btheta$ integrations (and using the constraints), we
obtain \beq L= 2|\Dt z|^{2}+\frac{i}{2}(\bpsi\cdot\Dt\psi-\Dt
\bar{\psi}\cdot\psi) -2ga. \label{susylag} \eeq  Substituting Eq.
(\ref{aeq}) into Eq. (\ref{susylag}), we can express the $N=2$
supersymmetric Lagrangian in the following form, \beq L= 2|\dot{z}
-({\bar z}\cdot \dot{z} )
z|^{2}+\frac{i}{2}(\bpsi\cdot\dot{\psi}-\dot{\bpsi}\cdot\psi)
-\frac{i}{2}(\bar{z}\cdot\dot{z}-\dot{\ol{z}}\cdot z)\bpsi\cdot\psi
+ig(\bar{z}\cdot\dot{z}-\dot{\ol{z}}\cdot z-i\bpsi\cdot\psi).
\label{susylagcom}\eeq The fact that the Lagrangian is
supersymmetric should be clear for it was written in terms of
superfields. At the component field level it can be most easily
verified using Eq. (\ref{susylag}).  It is interesting to note that
the supersymmetric magnetic monopole interaction term in Eq.
(\ref{susylagcom}) is given by the field $a$ itself. In this model
$U(1)$ gauge transformation of the monopole potential is realized by
the local $U(1)$ symmetry which is responsible for the reduction
from $S^{3}\rightarrow S^{2}$.  In Appendix A, we give the full
equations of motion.

\section{Dirac quantization and operator ordering}
\setcounter{equation}{0}
\renewcommand{\theequation}{\arabic{section}.\arabic{equation}}

In this section, we perform the canonical quantization of the
system.  We define the momenta $p$ and $\bar{p}$ conjugate to the
fields $z$ and $\bar{z}$, respectively by\footnote{Here, we have
chosen a specific ordering of the quantum mechanical operators $p$
and $\bar{p}$. Especially we take $\Dt z= \dot{z}-iza$ and its
conjugate $\Dt\bar{z}=\dot{\ol{z}}+ia\bar{z}$.}\beq
p=2\Dt\bar{z}+\frac{i}{2}(\bpsi\cdot\psi+2g)\bar{z},~~~ \bar{p}=2\Dt
z-\frac{i}{2}(\bpsi\cdot\psi+2g)z. \eeq  The Hamiltonian is obtained
as  \beq H=2|\Dt
z|^{2}-\frac{1}{2}(\bpsi\cdot\psi)^{2}-g\bpsi\cdot\psi,
\label{susyham} \eeq supplemented by the following four second class
constraints \beq C_{1}=\bar{z}\cdot z-1,~~~C_{2}=p\cdot
z+\bar{z}\cdot\bar{p},~~~ C_{3}=\bar{z}\cdot\psi,~~~C_{4}=\bpsi\cdot
z. \label{secconst} \eeq There is also a first class constraint
given by \beq C_{0}=-i(\bar{z}\cdot\bar{p}-p\cdot
z)-\bpsi\cdot\psi+2g, \label{firconst}\eeq which generates the
$U(1)$ transformation. We first define the Poisson brackets via \beq
\{z_{i},p_{j}\}=\{\bar{z}_{i},\bar{p}_{j}\}=\delta_{ij},~~~\{\bpsi_{i},\psi_{j}\}_+=-i\delta_{ij},
\eeq with the remaining brackets being zero.  We use the Dirac
brackets given by \beq
\{A,B\}_{D}=\{A,B\}-\{A,C_{a}\}\Theta^{ab}\{C_{b},B\}, \eeq where
$\Theta^{ab}$ is the inverse matrix of
$\Theta_{ab}=\{C_{a},C_{b}\}$. After some computation and quantizing
the Dirac brackets by replacing $\{A,B\}_{D}\rightarrow -i[A,B]$, we
obtain \beq
\begin{array}{ll}
\vspace{0.2cm}
\left[p_{i},z_{j}\right]=-i\delta_{ij}+\frac{i}{2}\bar{z}_{i}z_{j},
&\left[p_{i},\bar{z}_{j}\right]=\frac{i}{2}\bar{z}_{i}\bar{z}_{j},\\
\vspace{0.2cm}
\left[p_{i},p_{j}\right]=\frac{i}{2}(p_{i}\bar{z}_{j}-p_{j}\bar{z}_{i}),
&\left[\bar{p}_{i},p_{j}\right]=\frac{i}{2}(\bar{z}_{j}\bar{p}_{i}-z_{i}p_{j})
-\alpha\psi_{i}\bpsi_{j}+\beta\bpsi_{j}\psi_{i},\\
\left[\bpsi_{i},\psi_{j}\right]_+=\delta_{ij}-\bar{z}_{i}z_{j},
&\left[p_{i},\bpsi_{j}\right]=i\bpsi_{i}\bar{z}_{j}, \label{comm}
\end{array}
\eeq with $\alpha+\beta=1$. The above brackets are supplemented by
their Hermitian conjugates, and remaining commutators are zero.
Note that the brackets in the first and third lines of Eq. (\ref{comm}) have no
operator ordering ambiguity.  In the second line, the ordering of the
first bracket is fixed by the anti-symmetry property, while ordering
in the second bracket is chosen by the condition that the
variables $(z_{i},\bar{z}_{i},\psi_{i},\bpsi_{i},
p_{i},\bar{p}_{i})$ commute with the second class constraint,
$C_{2}$, ordered as $p\cdot z+\bar{z}\cdot\bar{p}=0$. Similar
ordering choice appeared before in the bosonic $CP(1)$ model
\cite{han}. Note that this does not fix the operator ordering
completely in the fermionic case, and we still have undetermined
$\alpha$ and $\beta$ in Eq. (\ref{comm}).

We then compute the Noether charge associated with phase symmetry of
the fermionic variables \beq N_F=\bpsi\cdot\psi, \eeq which turns
out to be the fermion number operator. In fact, using the
constraints one can derive the following result. \beq
(\bpsi\cdot\psi)^2=\bpsi\cdot\psi,\eeq therefore $N_F=0$ or $1$.
 The supersymmetry charges are
given by \beq Q=p\cdot\psi,~~~\bar{Q}=\bpsi\cdot\bar{p}.\eeq Note
that the supercharges have no ordering ambiguity. One can easily
check that $[\bpsi\cdot\psi,Q]=-Q$,
$[\bpsi\cdot\psi,\bar{Q}]=\bar{Q}$. Thus, $Q$ and $\bar{Q}$ play the
role of lowering and raising operator of the fermion number.

The global SU(2) rotations are generated by \beq \left(
\begin{array}{cc}
z_{1}\\
z_{2}
\end{array}
\right)\rightarrow e^{-\frac{i}{2}w^{a}\sigma^{a}} \left(
\begin{array}{cc}
z_{1}\\
z_{2}
\end{array}
\right), \eeq whose operator-ordered conserved charge is given by
\beq
K_{a}=\frac{i}{2}\bar{z}\sigma_{a}\bar{p}-\frac{i}{2}p\sigma_{a}z
+\gamma\bar{z}\sigma_{a}z+\frac{1}{2}\bpsi\sigma_{a}\psi. \eeq Here
we have added the third term associated
with the operator ordering ambiguity. After some computation, we
find that $K_{a}$'s  generate the SU(2) algebra \beq
[K_{a},K_{b}]=i\epsilon_{abc}K_{c},\label{angalg}\eeq provided the
following conditions are satisfied \beq
\alpha=\frac{1}{2}(1+4\gamma),~\beta=\frac{1}{2}(1-4\gamma).\eeq
Note that the angular momentum algebra Eq. (\ref{angalg}) is
satisfied independent of the value $\gamma$. In Appendix B, we give
various commutation relations of $K_a$'s with other operators and
express them in terms of  space unit vector $\vec x=\bar z
\vec\sigma z$.

\section{Energy spectrum and supersymmetry breaking}
\setcounter{equation}{0}
\renewcommand{\theequation}{\arabic{section}.\arabic{equation}}

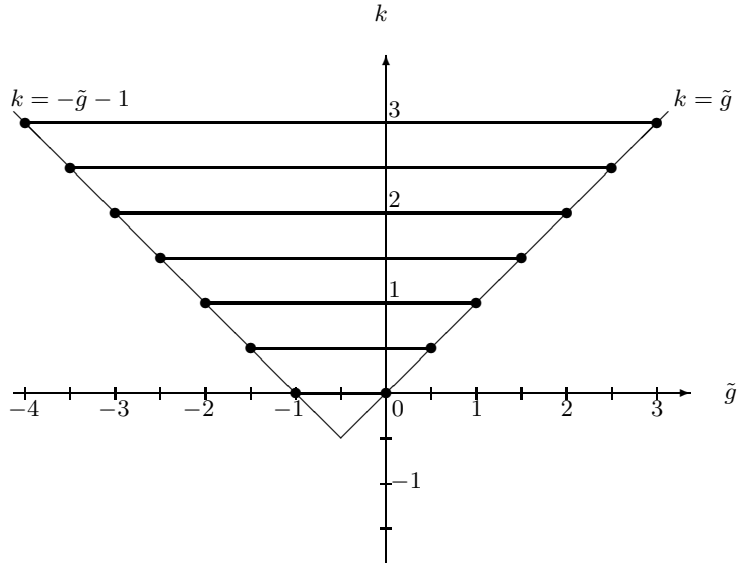
\begin{figure}[t]
\setlength{\unitlength}{1.5cm}
\begin{center}
\begin{picture}(6.5,6.5)(0,0)
\put(7,1.95){$\tilde{g}$} \put(3.9,5.3){$k$}
\put(6.55,4.55){$k=\tilde{g}$} \put(0.67,4.55){$k=-\tilde{g}-1$}
\put(4.04,1.16){$-1$} \put(4.02,2.85){1} \put(4.02,3.65){2}
\put(4.02,4.45){3} \put(4.75,1.8){1} \put(5.55,1.8){2}
\put(6.35,1.8){3} \put(4.05,1.8){0} \put(3.0,1.8){$-1$}
\put(2.2,1.8){$-2$} \put(1.45,1.8){$-3$} \put(0.65,1.8){$-4$}
\put(3.95,0.8){\line(1,0){0.1}} \put(3.95,1.2){\line(1,0){0.1}}
\put(3.95,1.6){\line(1,0){0.1}} \put(4.4,1.95){\line(0,1){0.1}}
\put(4.8,1.95){\line(0,1){0.1}} \put(5.2,1.95){\line(0,1){0.1}}
\put(5.6,1.95){\line(0,1){0.1}} \put(6.0,1.95){\line(0,1){0.1}}
\put(6.4,1.95){\line(0,1){0.1}} \put(1.2,1.95){\line(0,1){0.1}}
\put(2.0,1.95){\line(0,1){0.1}} \put(2.8,1.95){\line(0,1){0.1}}
\put(3.2,1.95){\line(0,1){0.1}} \put(3.6,1.95){\line(0,1){0.1}}
\put(2.4,1.95){\line(0,1){0.1}} \put(1.6,1.95){\line(0,1){0.1}}
\put(0.8,1.95){\line(0,1){0.1}} \put(4.8,1.95){\line(0,1){0.1}}
\put(0.7,2){\vector(1,0){6}} \put(4,0.5){\vector(0,1){4.5}}
\thicklines \put(3.2,2){\line(1,0){0.8}} \put(4,2){\circle*{0.1}}
\put(3.2,2){\circle*{0.1}} \put(2.8,2.4){\line(1,0){1.6}}
\put(2.8,2.4){\circle*{0.1}} \put(4.4,2.4){\circle*{0.1}}
\put(2.4,2.8){\line(1,0){2.4}} \put(2.4,2.8){\circle*{0.1}}
\put(4.8,2.8){\circle*{0.1}} \put(2,3.2){\line(1,0){3.2}}
\put(2,3.2){\circle*{0.1}} \put(5.2,3.2){\circle*{0.1}}
\put(1.6,3.6){\line(1,0){4.0}} \put(1.6,3.6){\circle*{0.1}}
\put(5.6,3.6){\circle*{0.1}} \put(1.2,4){\line(1,0){4.8}}
\put(1.2,4){\circle*{0.1}} \put(6,4){\circle*{0.1}}
\put(0.8,4.4){\line(1,0){5.6}} \put(0.8,4.4){\circle*{0.1}}
\put(6.4,4.4){\circle*{0.1}} \thinlines
\put(3.6,1.6){\line(1,1){2.9}} \put(3.6,1.6){\line(-1,1){2.9}}
\end{picture}
\end{center}
\caption{Diagram for $k$ versus $\tilde{g}$.}
\end{figure}

The quantum mechanical Hamiltonian is defined by \bea H_q&\equiv&
\frac{1}{2}[Q, \bar Q]_+\nn\\
&=&\frac{1}{2}p\cdot\bar p-\frac{1}{2}g^2-\frac{1}{2}(g+2\gamma)
\bpsi\cdot\psi-\frac{1}{8}(\bpsi\cdot\psi)^{2}, \label{normham} \eea
which differs from the classical Hamiltonian Eq. (\ref{susyham}) by
operator ordering.  This Hamiltonian can be expressed in terms of
$K_{a}$ operators as \beq
H_q=\frac{1}{2}\left[K^2-\left(g+\gamma\right)
\left(g+\gamma+1\right)\right].\eeq Thus, we obtain the energy
spectrum as follows \beq
E=\frac{1}{2}\left[k(k+1)-\tilde{g}(\tilde{g}+1)\right],\eeq where
$k=0, \frac{1}{2}, 1, \cdots$ is the angular quantum number
associated with $K_{a}$ operator and $\tilde{g}=g+\gamma$. Some
comments are in order at this point.  The energy $E$ must be
positive definite because of the first equation of (\ref{normham}).
Moreover, the spectrum is obtained by exploiting the rotational
invariance.  In the case of $\gamma=0$, the method of raising and
lowering operators can be used to construct the energy
eigenvalues~\cite{kim}. The Hamiltonian commutes with the fermion
number $N_{F}$ and $K_{a}$'s and thus the spectrum has a
$2(2k+1)$-fold degeneracy.

We observe that $k$ must satisfy the following inequality due to
the positive definiteness of the energy spectrum \beq k\ge
\vert\tilde{g}+\frac{1}{2}\vert-\frac{1}{2}. \eeq In Fig. 1, we
give the diagram for $k$ versus $\tilde g$. Each horizontal solid
line denotes the allowed values of $\tilde g$ for a given angular
momentum $k$. For a given value of $\tilde g$ in this range, the
energy spectrum is given by the vertical intersections with
$k=constant$ lines. The circular dots at the end of each
horizontal line represent the supersymmetric vacuum state, and for
these particular values of $\tilde g$, supersymmetry is unbroken.
Observe that there exists a reflection symmetry in the parameter
space of the monopole charge; a given value of $\tilde g$ yields
the same energy with $\tilde g^{\prime} = -\tilde g -1$. The
spectrum is symmetric with respect to $\tilde g=-\frac{1}{2}$
axis. Supersymmetry is spontaneously broken unless the minimum
values of $k$ reside on the lines $k=\tilde g$ or $k=-(\tilde
g+1)$ in Fig. 1. In other words, the breaking occurs unless the
parameter $\gamma$ is quantized, $\gamma=n/2$ for some integers
$n$.

Let us examine some cases. For symmetric ordering with the value of
$(\alpha,\beta,\gamma)=(\frac{1}{2},\frac{1}{2},0)$, the energy
spectrum is given by \beq E=\frac{1}{2}\left[k(k+1)-g(g+1)\right].
\label{egg}\eeq  With the substitution $k=n+g$ $(n=0,1,2,\cdots)$,
the above spectrum (\ref{egg}) agrees with the previous calculations
based on the method using the shape-invariance \cite{kim}. The
complete spectrum in Ref.~\cite{kim} also gives the same
multiplicities as our result. We observe that supersymmetry is
unbroken due to the Dirac quantization condition $g=n/2$ in this
case.
 For asymmetric ordering with value of
$(\alpha,\beta,\gamma)=(-\frac{1}{2},\frac{3}{2},-\frac{1}{2})$, we
have \beq E=\frac{1}{2}\left[k(k+1)-g^2+\frac{1}{4}\right].
\label{specta}\eeq Similar relation appeared in Ref. \cite{dejo}
where the Casimir invariant in the right hand side of Eq.
(\ref{specta}) is associated with the hidden supersymmetry which is
generated by the Killing-Yano tensor \cite{gibb} in the $N=1$
superspace approach.

\section{Summary and discussions}
\setcounter{equation}{0}
\renewcommand{\theequation}{\arabic{section}.\arabic{equation}}

In summary, we investigated $N=2$ supersymmetric quantum mechanics
of a charged particle on sphere in the background of magnetic
monopole. Our formulation  has a couple of novel aspects. First, we
introduced gauged chiral superfield which is annihilated by gauge
covariant superderivatives. These gauge covariant superderivatives
and their associated supercharges fulfill the usual relations of
supersymmetry in Eq. (\ref{nilpo}).  We also adopted $CP(1)$ model
approach which admits a compact $N=2$ superspace formulation of the
problem. Then, we carried out the Dirac quantization and computed
the exact quantum mechanical spectrum.  We found that the
spontaneous breaking of supersymmetry occurs unless the parameter
$\gamma$ is quantized.

 The spontaneous breaking of supersymmetry occurs in this system  for
generic values of $\gamma$.  Recall that the parameter $\tilde g$
which characterizes the breaking is composed of two factors; the
monopole charge $g$ and the parameter $\gamma$ representing the
effect of the operator ordering ambiguity. Even in the case without
monopole, spontaneous supersymmetry breaking occurs  except for the
case where $\gamma=n/2$. On the other hand, the monopole effect can
be dominant in the large $g$ case in which the background space
becomes fuzzy sphere \cite{hatsuda}. It would be interesting to
explore the connection between the fuzzy sphere and supersymmetry
further in the present framework.

\acknowledgments We would like to thank Choonkyu Lee and Jin-Ho Cho
for useful discussions.  STH would like to acknowledge financial
support in part from the Korea Science and Engineering Foundation
Grant R01-2000-00015.  The work of PO  was supported by the Korea
Research Foundation Grant R05-2004-000-10682-0.



\appendix
\section{Classical equations of motions}
\setcounter{equation}{0}
\renewcommand{\theequation}{A.\arabic{equation}}

In order to derive the classical equations of motion, we consider
the $N=2$ SUSY Lagrangian given by \bea L&=& 2|\dot{z} -({\bar
z}\cdot \dot{z} )
z|^{2}+\frac{i}{2}(\bpsi\cdot\dot{\psi}-\dot{\bpsi}\cdot\psi)
-\frac{i}{2}(\bar{z}\cdot\dot{z}-\dot{\ol{z}}\cdot z)\bpsi\cdot\psi
+ig(\bar{z}\cdot\dot{z}-\dot{\ol{z}}\cdot
z-i\bpsi\cdot\psi)\nonumber\\&+&N(\bar{z}\cdot
z-1)+\Lambda\bar{z}\cdot\psi+\bpsi\cdot z\bar{\Lambda},
\label{susylagconst}\eea where $N$, $\Lambda$ and $\bar{\Lambda}$
are the Lagrange multipliers associated with the second class
constraints derived from (\ref{susyconst}). Variations of the
Lagrangian (\ref{susylagconst}) over the variables $z$ and $\psi$
produce their equations of motion \bea
D_{t}p&=&-\frac{i}{4}(\bpsi\cdot\psi+2g)p-\frac{1}{8}(\bpsi\cdot\psi+2g)^{2}\bar{z}
+N\bar{z}-\bar{\Lambda}\bpsi,\nn\\
D_{t}\psi&=&\frac{i}{2}\psi(\bpsi\cdot\psi+2g)+iz\bar{\Lambda},
\label{eoms} \eea where the Lagrangian multipliers are given by
\bea N&=&D_{t}p\cdot z+\frac{i}{4}(\bpsi\cdot\psi+2g)p\cdot z
+\frac{1}{8}(\bpsi\cdot\psi+2g)^{2},\nn\\
\Lambda&=&-\frac{i}{2}\bpsi\cdot\bar{p},\nn\\
\bar{\Lambda}&=&\frac{i}{2}p\cdot\psi. \eea The equations of
motion for $\bar{z}$ and $\bpsi$ can be readily read off from the
Hermitian conjugates of the above corresponding equations.

\section{Quantum commutators}
\setcounter{equation}{0}
\renewcommand{\theequation}{B.\arabic{equation}}

Using the commutators Eq. (\ref{comm}), we find that \beq [K_a,
X]=-\frac{\sigma_a}{2}X,~~X=(z, \bar p, \psi),\eeq and \beq [K_a,
\bar X]=\bar X\frac{\sigma_a}{2},~~\bar X=(\bar z, p, \bar\psi).
\eeq Note that these relations hold independent of $\gamma$ and
confirm that $K_a$'s are indeed generators of rotations. We also
have \beq [K_a, Q]=[K_a, \bar Q]=[K_a, \bar\psi\cdot\psi]=0.\eeq In
terms of unit vector $\vec{x}=\bar z \vec{\sigma} z$, the angular
momentum generator $ \vec K\equiv K_a$ is given by \beq \vec K =
\vec x\times \dot {\vec x}
+\left(g+\gamma+1-\bpsi\cdot\psi\right)\vec x.\eeq Note that besides
$(\gamma+1)\vec x$ term, this expression is the same as the
well-known angular momentum in ${\bf R}^3$ in the bosonic case
\cite{jack}.

\end{document}